\documentstyle[seceq,epsf,twoside]{ptptex}
\notypesetlogo  
\title{Possible Evidence for a Chiral Axial-Vector State \\ 
in the $D$ Meson System}
\author{%
Kenji {\sc Yamada}$^*$, Muneyuki {\sc Ishida}$^{\P}$, 
Shin {\sc Ishida}$^{\dagger}$, Daiki {\sc Ito}$^{\dagger}$, \\
Toshihiko {\sc Komada}$^*$ and Hiroshi {\sc Tonooka}$^{\dagger}$ }
\inst{%
$^*$Department of Engineering Science, Junior College Funabashi Campus \\ 
Nihon University, Funabashi 274-8501, Japan\\
$^{\P}$
Department of Physics, Tokyo Institute of Technology, Tokyo 152-8551, Japan\\
$^{\dagger}$Atomic Energy Research Institute, 
College of Science and Technology\\
Nihon University, Tokyo 101-8308, Japan
}
\abst{%
We reanalyze the $ D^{*+} \pi^- $ mass spectrum from CLEO II by the VMW method in order to examine the 
existence of a chiral axial-vector state, which is predicted in a covariant level-classification scheme 
recently proposed, other than normal orbitally-excited $P$-wave states in the $D$ meson system. 
A result of the present analysis seems to suggest that there exists an extra axial-vector meson, 
in addition to the two normal ones, in a similar mass region.
}

\begin{document}
\maketitle

\setcounter{tocdepth}{4}

\section{Introduction}

In the constituent quark model, together with heavy quark symmetry, 
the lowest-lying positive parity excitations of heavy-light $Q \bar q$ meson systems are expected, 
in the limit $m_Q \to \infty$, to be two degenerate spin doublets with the total angular momentum 
$j_q=1/2$  and $3/2$ of the light quark, that is, four orbitally-excited states with $L=1$ labeled as
\begin{eqnarray}
^{j_q }L_J  = ^{1/2} \!\! P_0 ~,~ ^{1/2}P_1 \ \ \ \  \mbox{\rm for the } j_q = {\textstyle{{\rm 1} \over {\rm 2}}}  \ \ \mbox{\rm doublet,} \nonumber \\
            = ^{3/2} \!\! P_1 ~,~ ^{3/2}P_2 \ \ \ \  \mbox{\rm for the } j_q = {\textstyle{{\rm 3} \over {\rm 2}}}  \ \ \mbox{\rm doublet.} \nonumber 
\end{eqnarray}
In this limit heavy quark symmetry further requires that the $j_q=1/2$ states decay to $^{1/2}S_0 + \pi$ or $^{1/2}S_1 + \pi$ only in an $S$-wave, while the $j_q=3/2$ states decay only in a $D$-wave.  It is therefore expected that the decay widths of the $j_q=1/2$  and $3/2$ states are broad and narrow, respectively.

On the one hand, a covariant level-classification scheme of quark-antiquark meson systems 
has been proposed, which gives them a covariant quark representation with definite Lorentz and 
chiral transformation properties.\cite{rf1} In this scheme, assuming that chiral symmetry for 
the light-quark component in heavy-light meson systems is effective, the existence of extra scalar 
and axial-vector states is predicted, respectively, as chiral partners of the ground-state 
pseudoscalar and vector mesons.  These what we call chiral scalar and axial-vector mesons are 
distinguished from the above-mentioned $P$-wave states, since the chiral scalar state is an 
analogue of the $\sigma(400-600)$ meson, 
which is difficult to be interpreted as the $^3P_0$ state, as a chiral partner of the $\pi$ meson 
in the light-quark system.

In this report we present a possible evidence for the chiral axial-vector state, $D_1^\chi$, 
in the $D$ meson system.

\section{Reanalysis of the $ D^{*+} \pi^- $ mass spectrum  from CLEO II}
We reanalyze the $ D^{*+} \pi^- $ mass spectrum, published by CLEO Collaboration,\cite{rf2} 
by the VMW method in which the production amplitude is expressed by a sum of Breit-Wigner 
amplitudes for relevant resonances.

In the present analysis we take into account the four states $D_2^* (^{3/2}P_2)$,
$D_1 (^{3/2}P_1)$, $D_1^* (^{1/2}P_1)$ and $D_1^\chi$ which can decay to $D^* \pi$. 
Then, following the VMW method, the production amplitude is given by
$$
\left| A(s) \right|^2  = \left| r_1 e^{i\theta _1} \Delta _{D_1^\chi} (s) 
                              + r_2 e^{i\theta _2 } \Delta _{D_1^*} (s)  \right|^2  
                       + \left| r_3 e^{i\theta _3 } \Delta _{D_1 } (s)   \right|^2  
                       + \left| r_4 e^{i\theta _4 } \Delta _{D_2^*} (s)  \right|^2  ,
$$
\vspace{-1.1em}
$$ \Delta _R (s) = \frac{ - m_R \Gamma _R }{s - m_R^2  + im_R \Gamma _R },  $$
where $r_1 ,...,r_4$ and $\theta _1 ,...,\theta _4$ are the production couplings and phases of 
respective resonances, and we assumed that $D_1^\chi$ and $D_1^*$ decay only through an $S$-wave, 
while $D_1$ only through a $D$-wave.  The background $D^*\pi$ mass distribution is fit with a 
five-parameter threshould function given by
$$
BG = \alpha (\Delta M)^\beta  \exp \left[  - \gamma _1 \Delta M - \gamma _2 (\Delta M)^2  - \gamma _3 (\Delta M)^3  \right] , 
$$
\vspace{-1.1em}
$$
\Delta M = M(D^ *  \pi ) - m_{D^ *  }  - m_\pi ,
$$
where the parameters $\alpha$, $\beta$, $\gamma_1$,  $\gamma_2$ and $\gamma_3$ are fixed through the fit to the total $D^*\pi$ mass spectrum.

\begin{figure}[hbpt]
  \epsfxsize=14 cm
 \centerline{\epsffile{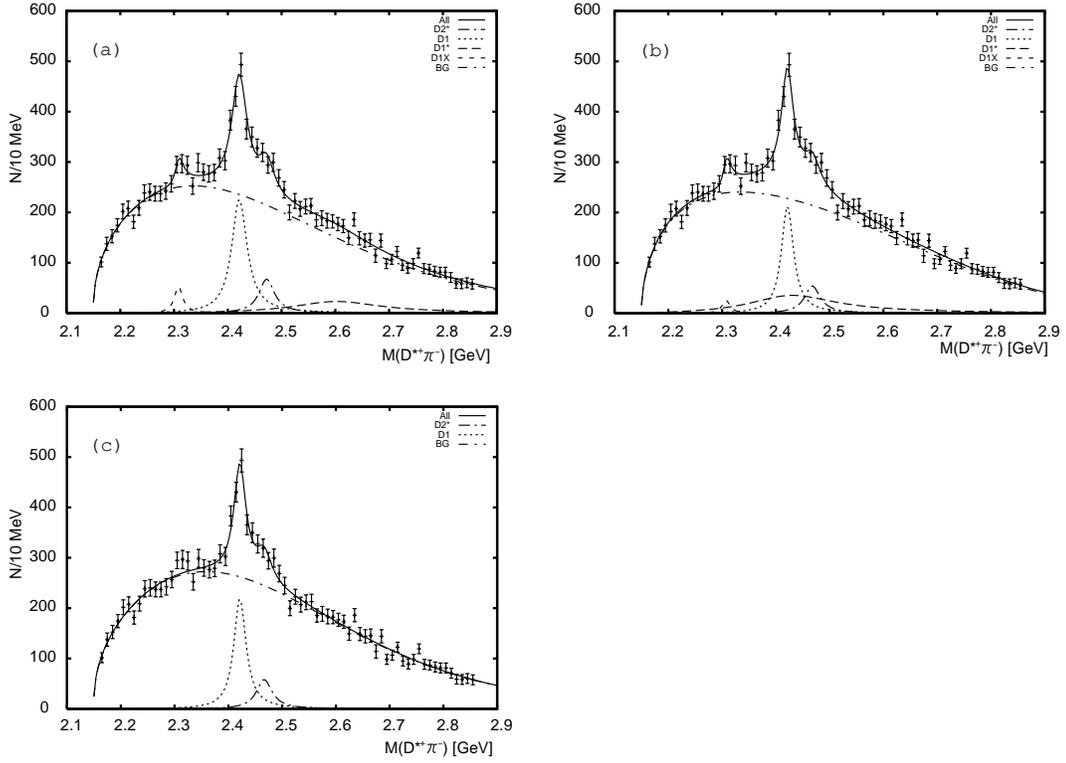}}
 \caption{
The results of the fits to the $ D^{*+} \pi^- $ mass spectrum with 
(a) high-mass $D_1^*$, (b) low-mass $D_1^*$, and (c) no $D_1^\chi$ and  $D_1^*$.}
  \label{fig:1}
\end{figure}

Using the above production amplitude and background, we fit the $D^*\pi$ mass spectrum in the 
following three cases:

\indent (a) High-mass $D_1^*$ with a mass of $2500 < m_{D_1^*} < 2600$ in MeV,\\
\indent (b) Low-mass $D_1^*$ with a mass of $2350 < m_{D_1^*} < 2500$ in MeV,\\
\indent (c) No $D_1^\chi$ and  $D_1^*$.\\
Here the case (c) corresponds to the original analysis by CLEO Collaboration, 
though the background parametrization is somewhat different. 
The results of fits are shown in Fig. \ref{fig:1} 
and obtained values of the resonance parameters are given in Table I.
In both the fits with high- and low-mass $D_1^*$, it is found that the mass and width of $D_1^\chi$  are 
$\approx$ 2310 MeV and $\approx$ 20 MeV, respectively, and those of $D_1$ and $D_2^*$ are similar to 
the values reported so far. For the mass and width of $D_1^*$ we obtain $\approx$ 2600 MeV and 
$\approx$ 200 MeV in the  high-mass fit, while $\approx$ 2420 MeV and $\approx$ 200 MeV in the low-mass fit.

In all the three cases of fits the ${\chi^2} / N_{\mbox{\rm {\footnotesize dof}}}$ is best for the high-mass $D_1^*$ fit, 
though they are not so different from each other. It would be worth while noting that the two cases of 
fits with $D_1^\chi$ seem to describe the data better than the fit without $D_1^\chi$ in the 
mass region 2.15$-$2.5 GeV, where there appears to be an excess of data events at the mass 2.31$-$2.33 GeV.

\begin{table}
\caption{Values of the resonance parameters and ${\chi^2} / N_{\mbox{\rm dof}}$ obtained from the respective fits.}
\begin{center}
\renewcommand{\arraystretch}{1.2}
\begin{tabular}{ccccccc}
\hline
\hline 
                        & \multicolumn{2}{c}{(a) Fit with high-mass $D_1^*$} 
                          & \multicolumn{2}{c}{(b) Fit with low-mass $D_1^*$} 
                            & \multicolumn{2}{c}{(c) Fit without $D_1^\chi$ and $D_1^*$ } \\ \cline{2-7}
              &  Mass   &  Width  &  Mass   &  Width  &   Mass  &  Width  \\
State         &  (MeV)  &  (MeV)  &  (MeV)  &  (MeV)  &  (MeV)  &  (MeV)  \\ \hline
$D_1^\chi$    &  2308   &   18.7  &   2307  &   17.4  &   ---   &   ---   \\
$D_1^*$       &  2596   &   199   &   2421  &   199   &   ---   &   ---   \\
$D_1$         &  2421   &   34.5  &   2421  &   27.0  &   2421  &   27.5  \\
$D_2^*$       &  2472   &   35.0  &   2468  &   35.0  &   2466  &   35.0  \\ \hline 
 ${\chi^2} / N_{\mbox{\rm dof}}$   & \multicolumn{2}{c}{ 57.7/52 } 
                          & \multicolumn{2}{c}{58.0/52} 
                            & \multicolumn{2}{c}{66.2/59} \\ \hline 
\end{tabular}
\end{center}
\end{table}

\section{Theoretical remarks on the results}

We consider the mass splitting and mixing of $P$-wave meson multiplets, based on the 
Breit-Fermi Hamiltonian with vector-gluon and long-range-scalar exchange, where we ignore the 
$P$-wave $D_1$ and $D_1^*$ states mixing with the chiral $D_1^\chi$ state. Taking a static potential 
due to single vector-gluon exchange to be $- 4\alpha _s / 3r$, the spin-dependent part of 
the Hamiltonian for $P$-wave states can be expressed, to first order $1/m_Q$, as
\begin{eqnarray}
\delta {\cal H} = C_q {\bf L} \cdot {\bf S}_q  + C_Q ({\bf L} \cdot {\bf S}_Q  + S_T ) , ~
S_T  = 3({\bf S}_q  \cdot {\bf \hat r})({\bf S}_Q  \cdot {\bf \hat r}) - {\bf S}_q  \cdot {\bf S}_Q \nonumber
\end{eqnarray}
with
\begin{eqnarray}
C_q  &=& \left( {\frac{1}{{2m_q^2 }} + \frac{1}{{m_q m_Q }}} \right)
       \left\langle {\frac{{4\alpha _s }}{{3r^3 }}} \right\rangle  
       - \frac{1}{{2m_q^2 }}\left\langle {\frac{1}{r}\frac{{dV_s }}{{dr}}} \right\rangle , \nonumber \\
C_Q  &=& \frac{1}{{m_q m_Q }}\left\langle {\frac{{4\alpha _s }}{{3r^3 }}} \right\rangle , \nonumber
\end{eqnarray}
where $V_s(r)$ is the static potential due to long-range-scalar exchange and the spin-spin interaction is 
neglected because of its contact nature. The Hamiltonian $\delta {\cal H}$ gives rise to the mass splittings 
among $P$-wave multiplets and the mixing between the $^{3/2}P_1$ and $^{1/2}P_1$ states. 
Treating $\delta {\cal H}$ as a first-order perturbation and using the mass values of $D_2^*$, $D_1$ and $D_1^*$ 
obtained in the high-mass $D_1^*$ fit, we find $\approx$ 2470 MeV for the mass of $D_0^*(^{1/2}P_0)$ and 
$\phi-\phi_{HQ} \approx -3.64^\circ$ for the deviation of the $D_1(^{3/2}P_1) - D_1^*(^{1/2}P_1)$ 
mixing angle from the heavy-quark-symmetry limit with the parameter values of the unperturbed mass 
$M_0  = 2490$ MeV common to all four states, $C_q  =  - 73.81$ MeV and $C_Q  = 47.15$ MeV, 
where we have chosen a solution with $C_Q  > 0$ in accord with the above definition of $C_Q$. 
For the low-mass $D_1^*$ case of fits there is no solution.

\section{Concluding remarks}

We have reanalyzed the $ D^{*+} \pi^- $ mass spectrum published by CLEO Collaboration and found a 
possible evidence for the chiral axial-vector meson $D_1^\chi$ with a mass and width of 
$\approx$ 2310 MeV and $\approx$ 20 MeV, respectively. We have also found the mass and width of 
$D_1^*$ to be $\approx$ 2600 MeV and $\approx$ 200 MeV, together with similar masses and widths of 
$D_1$ and $D_2^*$ to those reported so far, among which the spin-dependent Hamiltonian arising from 
one-vector-gluon and long-range-scalar exchange could account for the mass splittings. 
To confirm the existence of $D_1^\chi$ it goes without saying that further analyses, 
including other experimental data with high statistics, are necessary in a more precise way.

Furthermore, in establishing the covariant level-classification scheme of meson systems it is important 
to examine the existence of the chiral scalar meson $D_0^\chi$ as well as $B_0^\chi$ and $B_1^\chi$ 
in the $B$ meson system. An analysis of the $B \pi$ mass spectrum, to study the existence of $B_0^\chi$, 
is in progress and its preliminary result has been presented.\cite{rf3}


\begin{thebibliography}{99}
\bibitem{rf1}  Ishida, S., Ishida, M., and Maeda, T., Prog. Theor. Phys. {\bf 104}, 785-807 (2000).
\bibitem{rf2}  CLEO Collaboration, Phys. Letters B {\bf 331}, 236-244 (1994); {\bf 342}, 453(E) (1995).
\bibitem{rf3}  Ishida, M., and Ishida, S., these proceedings.
\end{thebibliography}
\end{document}